\documentclass[prd,twocolumn,showpacs,groupedaddress,superscriptaddress,amsmath,amssymb]{revtex4-2} 
 
\newcommand{\vect}[1]{\boldsymbol{#1}}

\def\nn{\nonumber}

\usepackage{color}	
\usepackage{graphicx}
\usepackage{bm}
	
\usepackage{slashed}

\usepackage[colorlinks=true, filecolor=blue, citecolor=blue,urlcolor=black]{hyperref}

\usepackage{enumitem}

\setenumerate{label=\arabic*)}

\date{\today}

\begin{document}

\title{Resonant probing  spin-0 and spin-2 dark matter mediators with fixed target experiments} 

\author{I.~V.~Voronchikhin}
\thanks{Corresponding author}
\email[\textbf{e-mail}: ]{i.v.voronchikhin@gmail.com}
\affiliation{ Tomsk Polytechnic University, 634050 Tomsk, Russia}

\author{D.~V.~Kirpichnikov}
\email[\textbf{e-mail}: ]{dmbrick@gmail.com}
\affiliation{Institute for Nuclear Research, 117312 Moscow, Russia}

\begin{abstract}
We discuss the mechanism to produce electron-specific dark matter mediators of spin-0 and spin-2
in the electron fixed target experiments such as NA64 and LDMX.  The secondary positrons induced  by the 
electromagnetic shower can produce the  mediators via annihilation on atomic electrons.
That mechanism, for some selected kinematics, results in the enhanced sensitivity with respect to 
the bounds derived by the bremsstrahlunglike  emission of the mediator in the specific parameter space. We 
derive the corresponding  experimental reach of the NA64 and LDMX.  
\end{abstract}

\maketitle

\section{Introduction}
The fundamental particle content of the Dark Matter (DM) 
can not be explained by the Standard Model (SM) even though it is associated with nearly 
$\simeq 85\%$ matter of the Universe~\cite{Planck:2015fie,Planck:2018vyg}. 
The indirect manifestations of the DM  are  related mainly to the galaxy rotation velocities, large-scale 
structures, the cosmic microwave background anisotropy, the gravitational lensing, etc~\cite{Bertone:2004pz,Bergstrom:2012fi,Gelmini:2015zpa}. 
However, the direct detection of the DM particles
remains one of the most significant challenges
of fundamental physics. 

 One can assume the thermal origin of the DM, implying the equilibrium between 
the DM and visible matter in the early Universe~\cite{Lee:1977ua}. 
In order to avoid the DM overproduction through thermal freeze-out, an idea of  
existence of the light massive mediator (MED) of DM  has been introduced 
\cite{Boehm:2002yz,Boehm:2003hm,Pospelov:2007mp,Arkani-Hamed:2008hhe,Krnjaic:2015mbs,Berlin:2018bsc}. 
For instance, the typical scenarios with dark boson mediators include spin-0, spin-1 and spin-2 particles 
such as the hidden Higgs boson~\cite{McDonald:1993ex,Burgess:2000yq,Wells:2008xg,Schabinger:2005ei,Bickendorf:2022buy,Boos:2022gtt,Sieber:2023nkq}, the dark 
photon~\cite{Holdom:1985ag,Okun:1982xi,Izaguirre:2015yja, Essig:2010xa,  Kahn:2014sra,Batell:2014mga,Batell:2017kty,Izaguirre:2013uxa,Kachanovich:2021eqa,Lyubovitskij:2022hna,Gorbunov:2022dgw,Claude:2022rho,Wang:2023wrx},  
and   dark  graviton 
\cite{lee:2014GMDM,Kang:2020-LGMDM,Bernal:2018qlk,Folgado:2019gie,Kang:2020yul,Dutra:2019xet,Clery:2022wib,Lee:2014caa,Gill:2023kyz,Wang:2019jtk,deGiorgi:2021xvm,deGiorgi:2022yha,Jodlowski:2023yne} respectively.  
Note that, the fermion DM portals have also recently been  discussed in the 
literature~\cite{Wojcik:2023ggt,Jueid:2023zxx,Kawamura:2022uft,Bai:2014osa,Kawamura:2020qxo,Xu:2023hkc}. 
The various mechanisms of DM thermalization involving mediators were studied by the authors of 
Refs.~\cite{Shakya:2015xnx,Lebedev:2023uzp,Poulin:2019omz,Dienes:2019krh,Bauer:2020nld,Poulin:2018kap}. For recent review 
of the probing these scenarios with  accelerator-based experiments,  see the 
Refs.~\cite{Feng:2022inv,Krnjaic:2022ozp,Gori:2022vri,Crivelli:2023pxa,Agrawal:2021dbo,Bondarenko:2023fex,Lanfranchi:2020kru,Lanfranchi:2020crw,Fox:2022tzz} and  references therein.

In this paper we focus on resonant probing the electron-specific spin-0 (spin-2) DM 
mediator, denoted by $\phi \, (G)$,  with electron fixed target experiments, such as 
NA64~\cite{Gninenko:2016kpg,NA64:2016oww,NA64:2017vtt,Gninenko:2017yus,NA64:2018lsq,Gninenko:2018ter,Gninenko:2019qiv,Banerjee:2019pds,NA64:2019auh,Dusaev:2020gxi,NA64:2020qwq,NA64:2020xxh,Bondi:2021nfp,NA64:2021xzo,NA64:2021aiq,NA64:2021acr,Andreev:2021fzd,NA64:2022rme,Arefyeva:2022eba,Zhevlakov:2022vio,Voronchikhin:2022rwc,Mongillo:2023hbs,Abdullahi:2023tyk}
and 
LDMX~\cite{Mans:2017vej,Berlin:2018bsc,Akesson:2022vza,Echenard:2019mdk,Moreno:2019tfm,Ankowski:2019mfd,Bryngemark:2021pfy,Schuster:2021mlr,Marsicano:2018glj}. 
 The typical production mechanisms
of spin-0 (spin-2) DM mediator   in the reaction of  high-energy electrons on a fixed target are associated with 
$\phi$ ($G$)-strahlung in nucleus scattering, $e^-N\to e^-N \phi \, (G)$, and resonant  annihilation
of secondary positrons on atomic electrons, $e^+ e^- \to \phi \, (G)$. In the present paper we consider mostly the 
invisible decay channels of mediators into pairs of specific DM particles, $\phi\, (G) \to\, \mbox{\rm DM} + \mbox{\rm DM}$.

It is worth mentioning that  all DM  benchmark scenarios of the present paper imply the light  $m_{\rm DM}$ to be in the range $1\, \mbox{MeV} - 1\, \mbox{GeV}$, such that 
these masses could be well within experimental reach of the current and forthcoming accelerator-based facilities.
Furthermore, these direct DM production limits are complementary to the DM relic-density parameter space~\cite{Berlin:2018bsc,Kang:2020-LGMDM}.

In addition, we note that for both  NA64 and LDMX experimental facilities, the resonant 
production of  $A'$ spin-1 DM mediator, $e^+ e^- \to A' \to \mbox{\rm DM} + \mbox{\rm DM}$, has been   studied  explicitly  in
Refs.~\cite{NA64:2022rme,Andreev:2021fzd,Schuster:2021mlr,Marsicano:2018krp,Celentano:2020vtu,Battaglieri:2021rwp,Marsicano:2018glj}. It was shown that including
the annihilation channel can provide the  improved bounds on thermal DM parameter space.  Therefore, the resonant 
probing  spin-0 and spin-2 DM mediators can extend the experimental reach of NA64e and LDMX, filling a gap in the 
existing literature~\cite{NA64:2021xzo,Voronchikhin:2022rwc,Berlin:2018bsc}.

This paper is organized as follows. 
In Sec.~\ref{SectBenchModels} we discuss the benchmark scenarios for the electron-specific DM 
mediators of spin-2 and spin-0.  
In Sec.~\ref{SecCrosSec} we  provide a description of existing expressions for the double differential cross section of 
 MED for the electron bremsstrahlunglike process in the case of Weizsäcker-Williams (WW) approach. 
In this section  we also derive explicitly the  resonant mediator production cross sections and its decay of widths  into DM.  In Sec.~\ref{sec:PosTrLgDis} we briefly discuss the main aspects of the positron track length distribution. 
In Sec.~\ref{ExperimentalBenchmark} we discuss the missing energy signal and overview the main benchmark parameters of the 
electron fixed target facilities. In Sec.~\ref{SecThresholds} we study the typical ranges of MED mass that yield the 
resonant enhancement of the MED parameter space. In Sec.~\ref{SectionExpectedReach} we obtain the 
experimental reach of  NA64 and LDMX for  spin-2 and spin-0 MED. We conclude in 
Sec.~\ref{SectionConclusion}.

\section{Benchmark scenarios
\label{SectBenchModels}}

\subsection{Tensor DM mediator}
Let us consider first the benchmark simplified coupling between the SM particles and the massive spin-2 
field $G_{\mu \nu}$ that is described by the following electron-specific effective Lagrangian 
\cite{lee:2014GMDM,Kang:2020-LGMDM},
\begin{align}
&    \mathcal{L}^{\rm G}_{\rm eff} 
\supset 
 -
    \frac{i c^{\rm G}_{ee}}{2\Lambda} G^{\mu \nu}
    \left(
            \overline{e} \gamma_{\mu} \overleftrightarrow{D}_{\nu} e
        -   \eta_{\mu \nu} \overline{e} \gamma_{\rho} \overleftrightarrow{D}^{\rho} e
    \right) \label{BechmarkLagrOftheModel1}  
\\& +  \frac{c^{\rm G}_{\gamma \gamma}}{\Lambda} G^{\mu \nu} 
    \left(
        \frac{1}{4} \eta_{\mu \nu} F_{\lambda \rho} F^{\lambda \rho} 
    +   F_{\mu \lambda} F^{\lambda}_{ \nu}
    \right) +
     \frac{c^{\rm G}_{\rm DM}}{\Lambda} G^{\mu \nu} T^{\rm DM}_{\mu \nu}, 
     \nonumber
\end{align}
where $e$ is the label of the SM electron, $F_{\mu\nu} = \partial_\mu A_\nu - \partial_\nu A_\mu$ is a 
stress tensor of the SM photon field $A_\mu$, $D_\mu = \partial_\mu - i e A_\mu$ is a covariant derivative of the $U(1)$ gauge field,  
and $\Lambda$ is the dimensional parameter for spin-2 interactions, that is associated with the scale of new physics;
$c_{ee}^{\rm G}$ and  $c_{\gamma \gamma}^{\rm G}$ are dimensionless couplings for  the  
electron and photon respectively. We choose the universal coupling 
$c^{\rm G}_{ee} = c^{\rm G}_{\gamma \gamma}$ throughout the paper,  in order to protect the unitarity of 
the scenario at low energies  (see e.~g.,~Refs.~\cite{Kang:2020-LGMDM} and references therein for detail). 
In addition, to be more concrete,  for the model with massive tensor mediator, $G_{\mu \nu}$, the DM 
candidates are chosen to be either a light Dirac fermion $\psi$ or hidden massive scalar 
$S$, such that the energy-momentum tensor $T^{\rm DM}_{\mu \nu}$ in Eq.~(\ref{BechmarkLagrOftheModel1}) is given, respectively, by~\cite{Kang:2020-LGMDM}    
\begin{align}
& T_{\mu \nu }^S = \partial_\mu S \partial_\nu S - \frac{1}{2}\eta_{\mu \nu} (\partial_\rho S )^2 - \frac{1}{2}\eta_{\mu \nu} m_S^2 S^2,  
\\
& T_{\mu \nu }^\psi = \frac{i}{4} \left[ \bar{\psi}(\gamma_\mu \partial_\nu + \gamma_\nu \partial_\mu) \psi - (\partial_\mu \bar{\psi} \gamma_\nu + \partial_\nu \bar{\psi} 
\gamma_\mu )  \right] 
\nonumber 
\\
& -\eta_{\mu \nu} (\bar{\psi} \gamma^\rho \partial_\rho \psi - m_{\psi} \bar{\psi} \psi ),
\end{align}
where $m_S$ and $m_\psi$ are the masses of $S$ and $\psi$, respectively. 
It is worth mentioning that the specific coupling constants $c^{\rm G}_{\rm DM} = (c^{\rm G}_{SS}, c^{\rm G}_{\psi\psi})$  in Eq.~(\ref{BechmarkLagrOftheModel1}) and the  set of masses $m_{\rm DM} = (m_{S}, m_{\psi})$ are assumed to be independent throughout the paper.

\subsection{Scalar mediator}
Let us consider now the electron-specific interaction with a light scalar DM mediator in the following form~\cite{Berlin:2018bsc} 
\[
    \mathcal{L}^{\phi}_{\rm eff}  \supset c^{\phi}_{e e} \,  \phi \, \overline{e} \, e, 
\]
where $c_{ee}^{\phi}$ is the dimensionless coupling of electron with 
$\phi$. The low-energy Lagrangian can arise through flavor-specific
five-dimensional effective operator~\cite{Batell:2017kty,Berlin:2018bsc,Forbes:2022bvo,Chen:2018vkr,Marsicano:2018vin} 
$$
 \mathcal{L}_{ {\rm dim} 5 } \supset  \sum_{i= e,\mu, \tau}  \left[ \frac{c_i}{\Lambda} \phi \, 
\overline{E}_L^i H e_R^i + \mbox{H.~c.} \right], 
$$
with $c_i$ being a Wilson coefficient for a flavor index~$i=(e,\mu,\tau)$, 
$H$, $E_L^i$ and $e_R^i$ are the SM Higgs doublet, lepton doublet and lepton singlet, respectively.
As emphasized in 
Ref.~\cite{Batell:2017kty,Berlin:2018bsc} one can choose the coupling of $\phi$  
predominantly to one flavor in order to avoid dangerous lepton flavor-violating
currents,  such that  $c_e\neq 0$ and $c_\mu \equiv c_\tau \equiv  0$. Note that various ultraviolet completions 
of that scenario have been suggested already, i.~e.,~in two-Higgs-doublet models involving extra-scalar singlet or  vectorlike quarks~\cite{Batell:2016ove,Batell:2017kty,Chen:2015vqy}.

Next, to be more specific, for the scenario with electron-philic scalar mediator 
$\phi$ we choose Majorana fermion as  well motivated DM  
candidate~\cite{Berlin:2018bsc}. The   effective Lagrangian reads as 
follows~\cite{Berlin:2018bsc,Nemevsek:2016enw}:
\begin{equation}
\mathcal{L}^{\phi}_{\rm Majorana \small } \supset - \frac{1}{2} c^\phi_{\chi\chi}  \phi \, \chi \, \chi + \frac{1}{2} m_{\chi} \chi \, \chi + \mbox{H.~c.}, 
\label{TypicalMajoranaCouplinToScalar}
\end{equation}
where $c^\phi_{\chi\chi}$ is the real dimensionless coupling, $\chi$ is a two component neutral 
Majorana DM and $m_\chi$ is its mass. The latter implies two on shell real degrees of freedom for $\chi$ as soon as  
$m_\chi \neq 0$.

\section{Cross section }\label{SecCrosSec}

\begin{figure*}[!htb]
\centering
\includegraphics[width=0.48\textwidth]{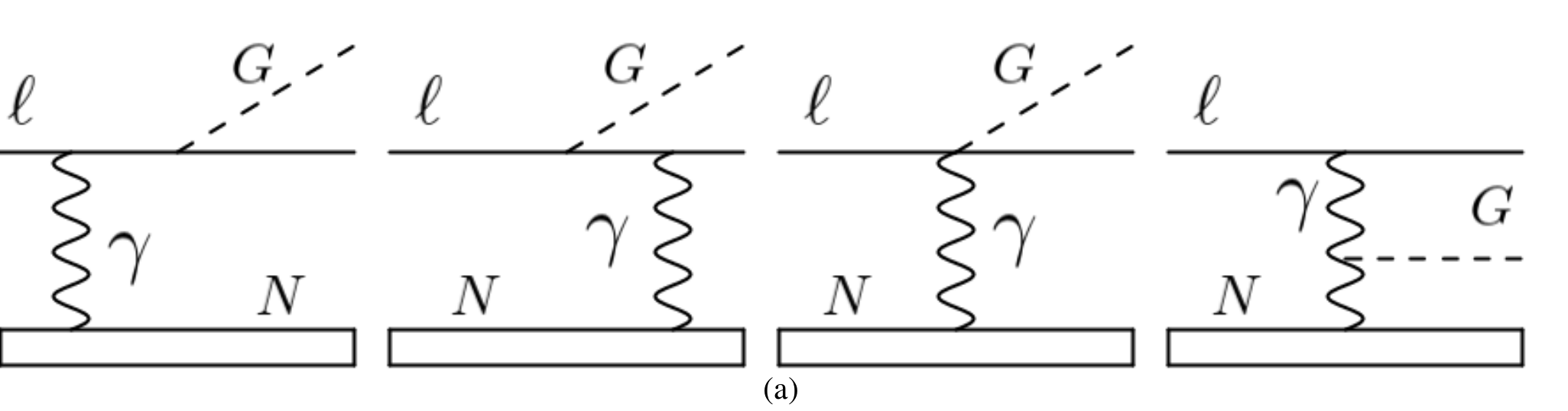}
\includegraphics[width=0.24\textwidth]{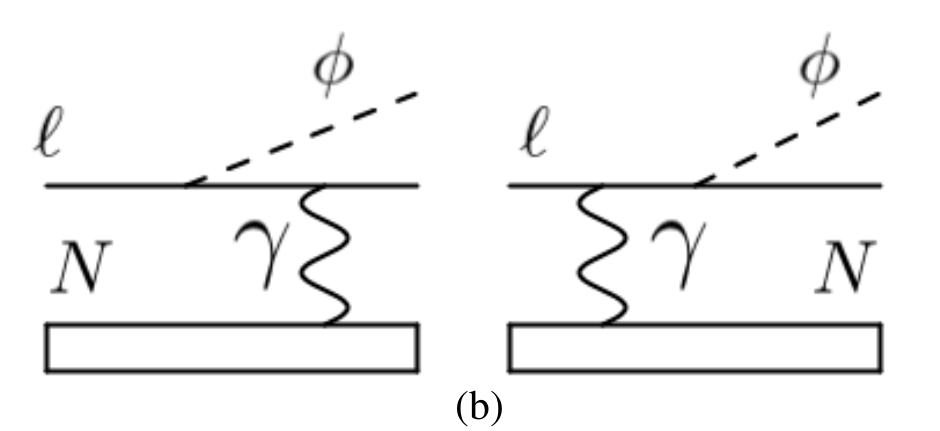}
\includegraphics[width=0.24\textwidth]{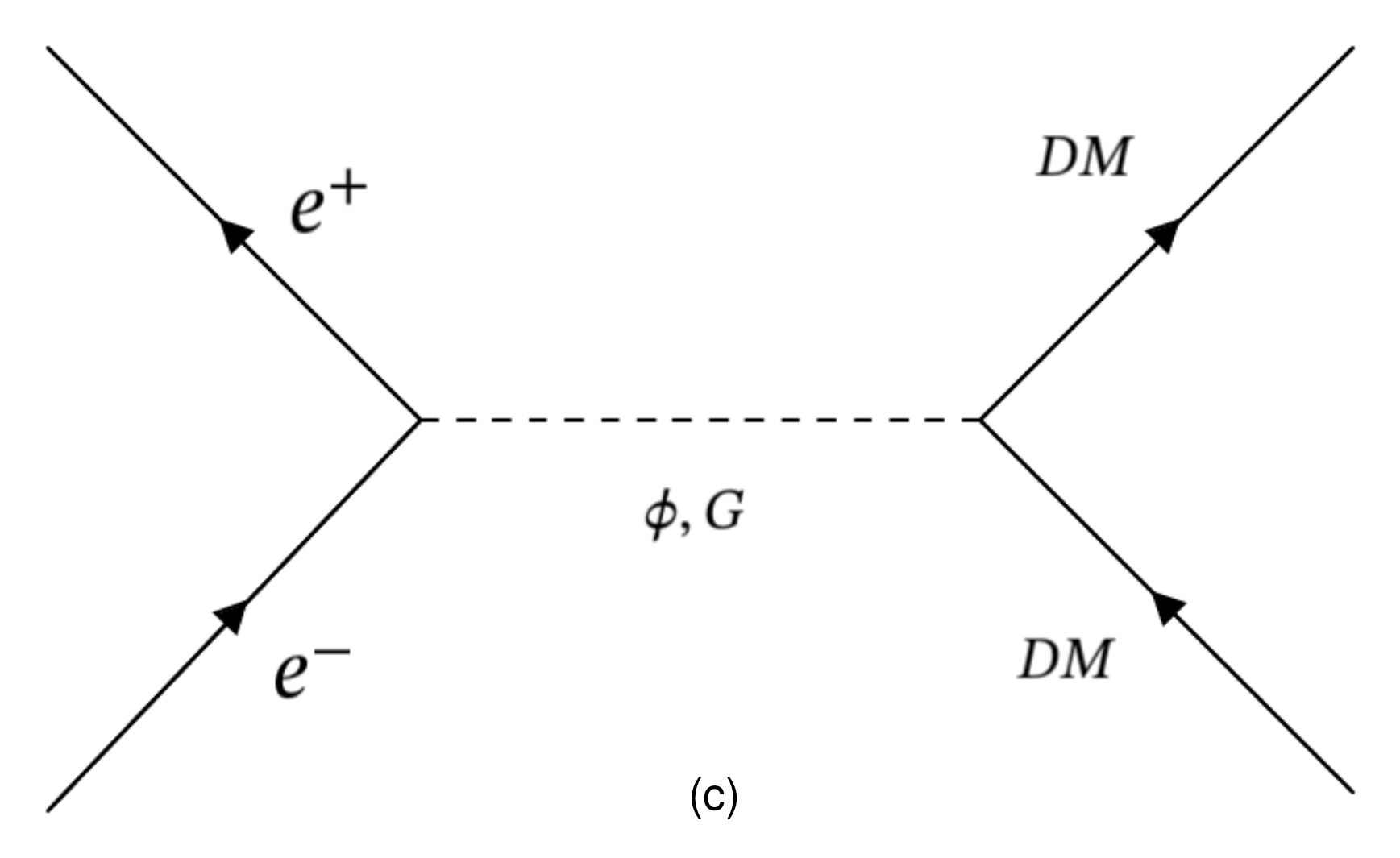}
\caption{Feynman diagrams describing production of $G$- (a) and $\phi$- (b)  
mediators via the bremsstrahlung missing energy process $e N \to e N G$ and $e N \to e N \phi$, respectively.
The diagram  (c) describes $s$-channel production of scalar and tensor DM mediators decaying resonantly 
into specific types of DM  particles: $e^+ e^- \to \phi \to \chi \chi$ and  $e^+ e^- \to G \to \psi \overline{\psi} 
(SS)$. 
}\label{fig:eNToeNGphi}
\end{figure*}

\subsection{Bremsstrahlunglike production of mediator}\label{bremsstrahlungLike}

In this subsection, we  consider the double differential cross section for a bremsstrahlung-like 
production of  MED of spin-0 and spin-2 by exploiting the WW approximation 
(see e.~g.,~Fig.~\ref{fig:eNToeNGphi} (a) and 1(b) for detail). This approach can be used for the approximation of 
 the exact tree-level  production cross sections at the level of $\lesssim 2\%$ for both hidden spin-0 and 
 spin-1  bosons~\cite{Liu:2016mqv,Liu:2017htz,Kirpichnikov:2021jev,Sieber:2021fue}.

The typical 
bremsstrahlung-like process in a case of electron primary beam reads as:
\begin{equation}\label{eq:BremsKinemat}
    e^-(p) + N(P_i) \rightarrow e^-(p') + N(P_f) + \text{MED}(k),
\end{equation}
where $p~=~(E_{0},\vect{p}) $, $p'~=~(E'_{0},\vect{p'})$ are the momenta of incoming and outgoing electrons, respectively, $k=(E_{MED}, \vect{k}) $ is the momentum of MED,  $P_i~=~(M,0)$ and  
$P_f~=~(P^0_f,\vect{P}_f)$ are the momenta of the initial and outgoing nucleus, respectively, $ P_i-P_f=q$, where  
$ q=(q_0,\vect{q})$ is the four-momentum of transfer to nucleus, we define $t \equiv - q^2$ throughout the paper. 
Also, the Mandelstam variables take the following form
\begin{equation}
    s_2 = ( p + q )^2, 
\quad 
    u_2 = ( p - k )^2, 
\quad
    t_2 = (p - p')^2.
    \label{MandelstamDef1}
\end{equation}
For the production of spin-2 MED in a process \eqref{eq:BremsKinemat}, the double differential cross
section for the approach $m_e / m_G \ll 1$ is given  by the following 
expression~\cite{Voronchikhin:2022rwc}: 
\begin{align}
&
    \left.\frac{d \sigma ( p + P_i \rightarrow  p' + P_f + k )}{dx d\cos(\theta_{G})}\right|_{WW}
=
    -
    \frac{\alpha \chi}{ \pi }
    \frac{E_{e^-}^2 x \beta_{G}}{1-x}
    \frac{1}{8 \pi s_2^2}
\nn \\
&
 \times    4 \pi  \alpha \frac{(c_{ee}^{\rm G})^2}{\Lambda^2}
    \frac{ [(t_2\! +\! u_2)^2\!\! +\!\! (u_2 \!- \!m_G^2)^2] [4 u_2 s_2 \!+ \! m_G^2 t_2] }
    {4 t_2 u_2 s_2},
    \label{eq:dsWWResG}
\end{align}
where $\alpha = e^2 / ( 4 \pi ) \simeq 1/137$ is the fine structure constant, $x = E_{G}/E_{0}$ is 
the energy fraction that spin-2 MED carries away, $\theta_G$ is the angle between the initial lepton direction and the momentum of the produced $G$-boson and 
$\beta_{G} = \sqrt{1 - m_{G}^2 / (x E_{0})^2}$ is the typical velocity of $G$-boson.   
The flux of virtual photon $\chi$ from nucleus is expressed through Tsai's elastic form-factor as 
follows~\cite{Bjorken:2009mm,Darme:2020sjf,Liang:2021kgw,Liu:2023bby}:
\begin{equation}
\chi = 
   Z^2 \int\limits^{t_{max}}_{t_{min}} \frac{t - t_{min}}{t^2} \left( \frac{t}{(t_a+t)}  \frac{1}{(1 + t/t_d)} \right)^2 dt,
   \label{ChiDefininition1}
\end{equation}
where $\sqrt{t_{a}} = 1/R_a$ is a momentum transfer associated with nucleus Coulomb field screening due to the atomic
electrons, with $R_a$ being a  typical  magnitude of the atomic radius $R_a = 111 Z^{-1/3}/m_e$, 
$\sqrt{t_{d}} = 1/R_n$ is the typical momentum associated 
with nuclear radius $R_n$, such that $R_n\simeq 1/\sqrt{d}$ and $d = 0.164 A^{-2/3}  \text{GeV}^2$.
We also note that each Mandelstam variable in Eqs.~(\ref{MandelstamDef1}) and (\ref{ChiDefininition1}) is a 
function of $E_{e-}$, $m_G$ $x$ and  $\theta_G$ in the WW approach, the explicit expressions for these functions
(i.~e.~ for $s_2$, $u_2$, $t_2$, $t_{min}$ and $t_{max}$) are given in  Ref.~\cite{Voronchikhin:2022rwc}.

Also, for a similar $\phi$-strahlung process with assuming $m_e / m_\phi \ll 1$ the differential cross section takes the following  form \cite{Liu:2016mqv}:
\begin{align}
&
    \left.\frac{d \sigma ( p + P_i \rightarrow  p' + P_f + k )}{dx d\cos(\theta_{\phi})}\right|_{WW}
=
    \frac{\alpha \chi}{ \pi }
    \frac{E_{e^-}^2 x \beta_{\phi}}{1-x}
    \frac{4 \pi \alpha (c_{ee}^\phi)^2}{8 \pi s_2^2}
\nn \\
& 
  \times  \left( \frac{x^2}{1-x} + 2 m_\phi^2\frac{u_2 x + m_\phi^2(1-x) }{u_2^2} \right),
    \label{eq:dsWWResPhi}
\end{align}
where $x = E_{\phi}/E_{0}$ is the energy fraction that is carried away by the spin-0 mediator,
 $\theta_\phi$ is the angle between the initial electron direction and the momentum of the
 produced $\phi$-boson and $\beta_{\phi} = \sqrt{1 - m_{\phi}^2 / (x E_{0})^2}$ is the typical velocity of spin-0 
 boson.

It is worth mentioning that for a calculation of flux of virtual photons one can use such form-factors as 
Helm or exponential~\cite{Chen:2011xp,Lewin:1995rx,Dobrich:2015jyk,Freese:1987wu}.  However,
for the mass range of interest $m_{\rm MED} \lesssim 1\, \mbox{GeV}$,
the exploiting of these form-factors does not provide a sizable impact on the magnitude 
of the virtual photon  flux~\cite{Voronchikhin:2022rwc}.

\begin{figure}[!t]
\centering
\includegraphics[width=0.45\textwidth]{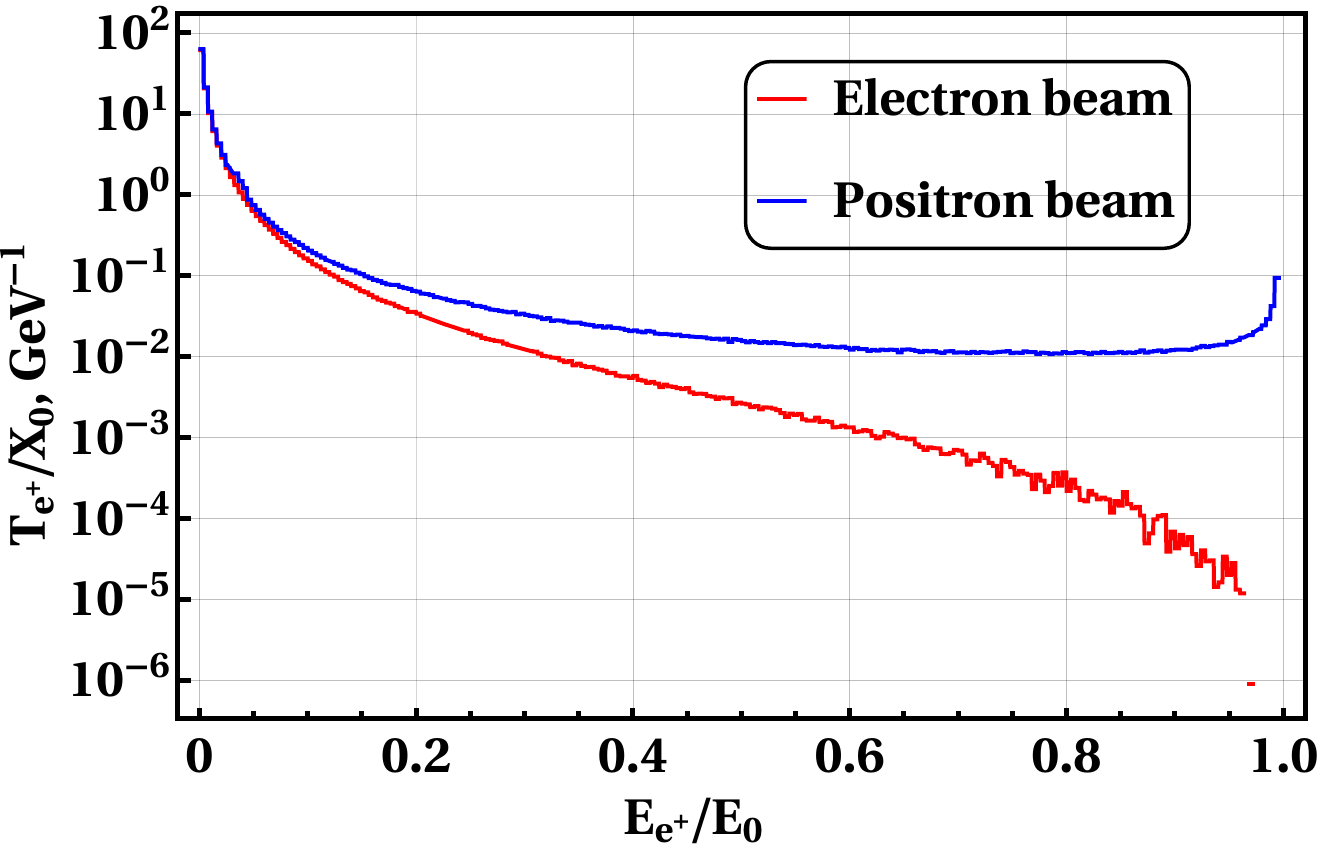}
\caption{The differential positron track-length distribution as function of ratio between the secondary positron energy $E_{e^+}$ and a primary beam energy $E_0$
of the electron (red line) or positron (blue line). 
}\label{fig:Tpos}
\end{figure}

 \subsection{The resonant production mechanism}
In this subsection, for the  benchmark  scenarios considered in Sec.~\ref{SectBenchModels},
we discuss  the  production cross section of the $\phi (G)$-boson that is  associated 
with  the annihilation of the secondary  positrons  with atomic electrons (see e.~g.,~Fig.~\ref{fig:eNToeNGphi}  (c)
for detail).

The first key ingredient for the  annihilation cross-section calculation is the partial decay width of hidden bosons. 
For the case of  the spin-2 boson the  decay widths into pairs of light scalars, dark fermions and 
a electron-positron  pair read, respectively,  as follows \cite{lee:2014GMDM}:
\begin{align}
    &     \Gamma_{G \to \psi \overline{\psi}}
=
    \frac{4 \pi \alpha_\psi m_G^3 }{160 \pi} 
       \left(\! 1 + \frac{8}{3}\frac{m_\psi^2}{m_G^2} \! \right)
    \left(\! 1 - 4 \frac{m_\psi^2}{m_G^2}\! \right)^{3/2},
\label{Gto2chiDecayWidth}    
\\ 
    &     \Gamma_{G \to SS}
=
    \frac{4 \pi \alpha_S m_G^3}{960 \pi}    
    \left( 1 - 4 \frac{m_S^2}{m_G^2} \right)^{5/2},
\label{Gto2SDecayWidth}    
\\
  &  \Gamma_{G \to e^+ e^-} \simeq 
    \frac{(c_{ee}^{\rm G}/\Lambda)^2 m_G^3 }{160 \pi} ,
\label{Gto2lDecayWidth}    
\end{align}
where we denote the dimensional dark-fine structure constants as $\alpha_\psi = (c^{\rm G}_{\psi\psi})^2 /(4 \pi \Lambda^2)$ and $\alpha_S = (c^{\rm G}_{SS})^2 /(4 \pi \Lambda^2)$, we also imply that $m_e/m_G\ll 1$. The resonant total cross section for spin-2 MED for the case of  Dirac  DM is:
\begin{multline}\label{eq:tot_eeToGchichi}
    \sigma_{e^-e^+ \to G \to \psi \overline{\psi}}
=
    \frac{4 \pi \alpha_\psi (c_{ee}^{\rm G})^2 }{256 \pi \Lambda^2}
    \frac{s^3}
         {(s - m_{G}^2)^2 + m_{G}^2 (\Gamma^{tot}_{G})^2}
\times \\ \times     \left( 1 +  8 m_{\psi}^2/(3 s) \right)
    \left( 1 - 4 m_\psi^2/ s\right)^{3/2},
\end{multline}
where $s = (p_{e_-} + p_{e_+})^2$ is Mandelstam variable, $p_{e_-}$ and  $p_{e_+}$ are the typical 4-momenta 
of the atomic electrons and secondary positrons, respectively.
 In the case of scalar DM the cross section takes following form:
\begin{multline}\label{eq:tot_eeToGToSS}
    \sigma_{e^-e^+ \to G \to S S}
\!\! =\!\!
    \frac{4 \pi  \alpha_S (c_{ee}^{\rm G})^2 }{1536 \pi \Lambda^2}
    \frac{s^3  \! \left(\! 1 \! - \! 4 m_S^2/s\right)^{5/2}}
         {(s \! - \! m_{G}^2)^2 \!  +\!  m_{G}^2 (\Gamma^{tot}_{G})^2}.
\end{multline}
For the calculation of both the decay width and cross section associated with spin-2 mediator we implement 
Feynman rules from Ref.~\cite{lee:2014GMDM} into the state-of-the-art FeynCalc 
package~\cite{Shtabovenko:2020gxv,Shtabovenko:2016sxi} for the Wolfram Mathematica routine~\cite{Mathematica}. 
In addition, we reproduce the well-known results for the massive graviton decay widths~\cite{lee:2014GMDM}.

Next, the widths of the decay of the scalar MED into  both the Majorana DM and electron-positron pair read, 
respectively as follows:
\begin{align} 
    &     \Gamma_{\phi \to \chi \chi}
=
   \frac{1}{2}\cdot
    \frac{4 \pi \alpha_\chi m_\phi }{8 \pi} 
    \left( 1 - 4 \frac{m_\chi^2}{m_\phi^2} \right)^{3/2},
\label{Sto2chiDecayWidth}    
\\ 
    &     \Gamma_{\phi \to e^+e^-}
\simeq 
    \frac{(c^{\rm \phi}_{ee})^2 m_\phi }{8 \pi} ,
\label{StoeeDecayWidth}    
\end{align}
where $\alpha_\chi = (c_{\chi \chi}^\phi )^2/(4 \pi)$ is a dimensionless dark-fine structure constant.

Finally, the resonant total cross section in the case of scalar MED and Majorana  DM reads as:
\begin{multline}\label{eq:tot_eeToSchichi} 
    \sigma_{e^+e^- \! \to \!  \phi  \! \to \!  \chi \chi}
\!\!=\!\!
  \frac{1}{2} \cdot 
    \frac{4 \pi \alpha_\chi (c^{\phi }_{ee})^2}{16 \pi }
    \frac{\!\! s\! 
        \left( \! 1 -\!  4 m_\chi^2/s \right)^{3/2}
        }{(s\! -\! m_{\phi}^2)^2 \! + \!  m_{\phi}^2 (\!\Gamma^{tot}_{\phi}\!)^2}.
\end{multline}

The derivation of the decay width~(\ref{Sto2chiDecayWidth}) and cross section~(\ref{eq:tot_eeToSchichi}) with Majorana 
particles in the final state is carried out in Appendix \ref{sc:MajorFerm}. It is important to emphasize that the total phase 
space for the 2-component Majorana spinors  should be  multiplied by the additional factor of $1/2$ in order to take into 
account the identical neutral   particles $\chi$ (see e.~g.~Ref.~\cite{Dreiner:2008tw} and references therein for 
detail).  Therefore, the well-known the result for the cross section with  Dirac fermions $\psi_D$ in the 
final state~\cite{Krnjaic:2015mbs} can be reproduced  from Eq.~(\ref{eq:tot_eeToSchichi}) if 
one omits its prefactor $1/2$. The latter cross-check can be also carried 
out  for  the  decay width of scalar into Majorana particles Eq.~(\ref{Sto2chiDecayWidth}).  
That implies the typical Lagrangian terms for the Dirac 4-components fermion  
$$\mathcal{L}^\phi_{\rm Dirac} \supset  c^{\phi}_{\psi_D \psi_D} \phi  \overline{\psi}_D \psi_D  + m_{\psi_D} \overline{\psi}_D \psi_D,$$
and the replacement 
$c_{\chi \chi }^\phi \to c^{\phi}_{\psi_D \psi_D}$  and $ m_\chi \to m_{\psi_D}$ in 
Eqs.~(\ref{Sto2chiDecayWidth}) and~(\ref{eq:tot_eeToSchichi})
(compare it with~Eq.~(\ref{TypicalMajoranaCouplinToScalar}) for clarification).

In the present study, we focus on the invisible decay mode of the mediators, which means that 
$m_{\rm MED} \gtrsim 2 m_{\text{\rm DM}}$ and $\Gamma_{{\rm MED} \to e^+e^-}~\ll~\Gamma_{\rm MED \to DM\;DM}$ throughout the paper.  This implies three benchmark decay widths $$\Gamma^{tot}_{G} \simeq  \Gamma_{G \to \psi \overline{\psi}}, \quad \Gamma^{tot}_{G} \simeq  \Gamma_{G \to SS }, \quad  \Gamma^{tot}_{\phi} \simeq  \Gamma_{\phi \to \chi \chi },$$ for the analysis of the electron missing energies signatures associated with accelerator based experiments NA64 and LDMX.  
Therefore, it leads to the rapid decay of
the mediator into a pair of DM after its production.

\section{Positron track-length distribution
\label{sec:PosTrLgDis}}

In this section we briefly discuss the distribution of positrons in a target  due to 
the electromagnetic shower development from the incoming  primary electron (positron) beams. 

The analytical approximations for the typical positron track-length distribution 
$T(E_{e^+})$ were studied in 
detail~\cite{Bethe:1934za,Carlson:1937zz,Landau:1938qvy,Marsicano:2018krp,Tsai:1966js}. 
For the thick target  it was shown that $T(E_{e^+})$ depends, at first order, on a specific 
type of target material through the multiplicative factor $T(E_{e^+}) \propto X_0$, that 
corresponds to the radiation length ($X_0$ is a typical distance over which a high-energy electron (positron)
loses all but  $1/e$ of its energy due to the bremsstrahlung, 
$e^{\pm}N \to e^{\pm} N \gamma$, where $e\simeq 2.71828 $ is a Euler's number).
Moreover,   $T(E_{e^+})$ depends also on the ratio $E_{e^+}/E_0$, where
$E_{0}$ is the energy of the primary impinging particle that initiates electromagnetic shower development
in the thick  target (see e.~g.~Fig.~\ref{fig:Tpos} for detail), 
so that one can exploit the typical 
distribution  shown in Fig.~\ref{fig:Tpos} for the specific type of electron fixed target experiment that is 
characterized  by energy of primary beam $E_0$ and target material~$X_0$. The distribution in Fig.~\ref{fig:Tpos} is 
adapted~\footnote{We would like to thank Andrea Celentano for sharing the  code and 
numerical data of Fig.~\ref{fig:Tpos}.} from Refs.~\cite{NA64:2022rme,Andreev:2021fzd}, that 
implies numerical Monte Carlo simulations for the electromagnetic (EM) 
shower development in GEANT4~\cite{GEANT4:2002zbu}. 

To conclude this section we note that a  positron track length depends also on the typical angles between the primary 
beam direction and momentum of the secondary positrons. 
However, this dependence impacts on the electromagnetic shower development  at the level
of $\lesssim  \mathcal{O} (1 \%)$ and hence the angular effects can be safely neglected in the 
estimates~\cite{Marsicano:2018krp}.

\section{Missing energy signal \label{ExperimentalBenchmark}}

In this section we  discuss the electron  missing energy signatures and the typical  parameters of the NA64 and LDMX experiments for probing DM mediators. It is worth mentioning that both NA64$e$ and LDMX experiments have background suppression at the level of 
$\lesssim~ \mathcal{O}(10^{-13}-10^{-12})$. 

\begin{figure*}[]
\centering
\includegraphics[width=1.0\textwidth]{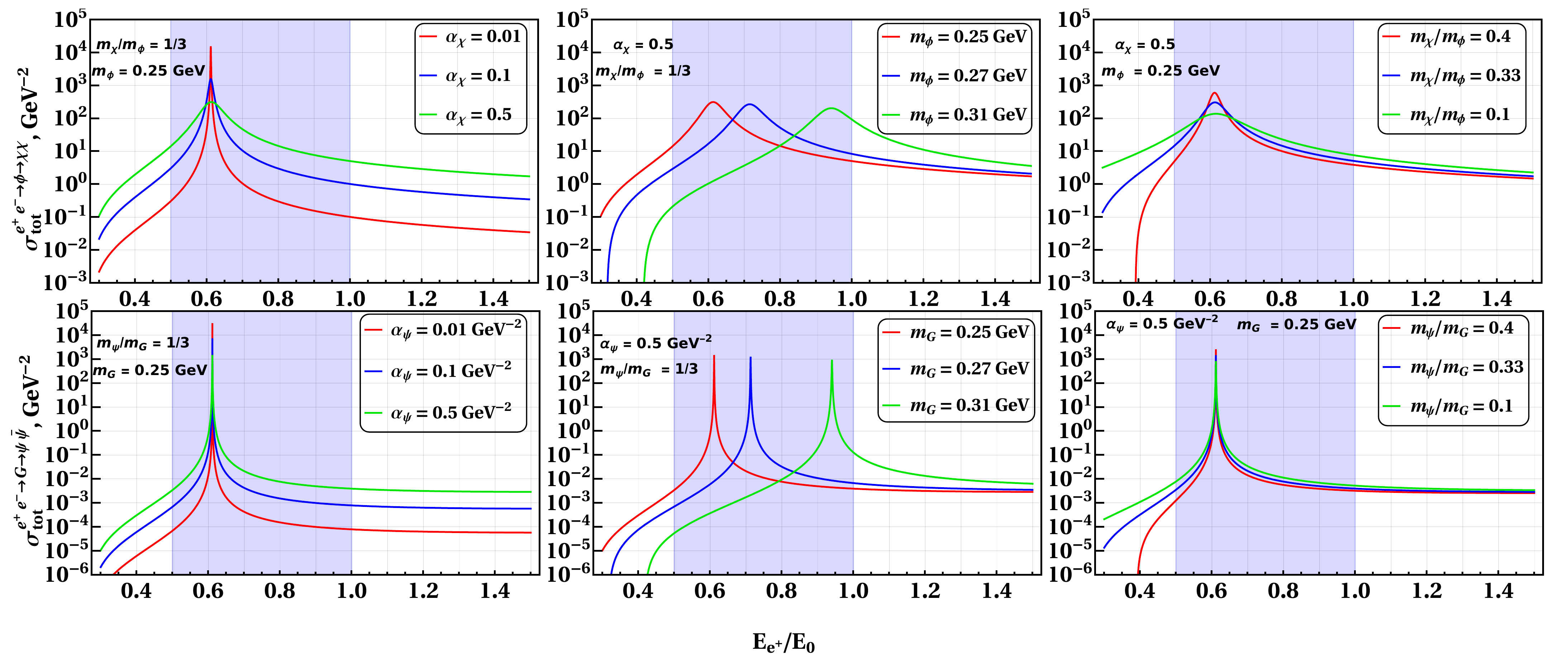}
\caption{The resonant total cross section for a scalar mediator  \eqref{eq:tot_eeToSchichi}  (upper panels) 
and for a tensor mediator \eqref{eq:tot_eeToGchichi}  (bottom panels) as the function of ratio between 
positron energy and energy primary beam for the set of coupling constants and masses.  
For the spin-2 and spin-0 cross sections we set
$c^{\rm G}_{ee}/\Lambda = 1  \;\mbox{GeV}^{-1}$ and $c^{\phi}_{ee} = 1$ respectively. 
The blue shaded region corresponds to the typical parameters space associated with the missing energy cuts 
$0.5 \lesssim E_{e^+}/E_0 \lesssim 1$
of NA64e for  $E_0 = 100 \;\mbox{GeV}$.
}\label{fig:tot_cross_sec_fermion}
\end{figure*}

We estimate the number of MEDs produced due to the 
bremsstrahlung for fixed target facilities as follows 
 \begin{equation}\label{eq:NradGrav}
N^{\rm brem. }_{\rm MED} \simeq \mbox{EOT}\cdot \frac{\rho N_A}{A} L_T \int\limits^{x_{max}}_{x_{min}}
dx \frac{d \sigma_{2\to3}(E_0)}{dx}\eta_{\rm MED}^{\rm brem.},
 \end{equation}
where   $L_T$ is effective interaction length of the electron in the target, 
$\mbox{EOT}$ is number of  electrons accumulated on target, $\rho$ is target density, $N_A$ is Avogadro's number,  
$A$ is the atomic weight number, $Z$ is the atomic number, $\eta_{\rm MED}^{\rm brem.}$ 
is a typical efficiency for the bremsstrahlung emission of the MED, $d\sigma_{2\to3}/dx$ is the 
differential cross section of the electron missing energy process $e N \to e N \phi ( G )$, 
$E_0 \equiv E_{beam}$ is the initial energy of electron beam,   $x_{min}$ and $x_{max}$
are the minimal and maximal fraction of missing energy, respectively
for the    experimental setup, $x\equiv E_{miss}/E_0$, where $E_{miss} \equiv  
E_{\rm MED}$. 
The $x_{min} \lesssim x \lesssim x_{max}$ cut is determined by specific fixed–target facility.

The typical number of hidden bosons produced due to the annihilation is estimated to 
be~\cite{Marsicano:2018krp,NA64:2022rme}:
\begin{equation}\label{eq:constrCoulConst}
    N^{\rm ann. }_{\mbox{\scriptsize MED}}\! \simeq \! \mbox{EOT} \frac{\rho N_A \! Z \! L_T }{A}\! \! \!\!\!\! \int\limits_{E_{e^+}^{cut}}^{E_{e^+}^{max}} \!
    \!\!\!\!\!  dE_{e^+} \! \sigma_{tot} (\!E_{e^+}\!)\! T(\!E_{e^+}\!)\eta_{\rm MED}^{\rm ann.},
\end{equation}
where  $\sigma_{tot} (E_{e^+})$ is the resonant total cross 
section of the electron-positron annihilation into DM, $\eta_{\rm MED}^{ann.}$ is a typical efficiency associated with MED production via the resonant channel (we conservatively imply throughout the paper that signals for both positron and electron beam modes have the same efficiency $\eta_{\rm MED}^{\rm ann.}$), $E_{e^+}$ is the energy of secondary positrons,   
$E_{e^+}^{cut}=E_{0} x_{min}$ and $E_{e^+}^{max} = E_{0}$ are the minimal and maximal energies of 
secondary positrons in the EW shower respectively. 

{\it NA64e: } the NA64e is the  fixed target experiment located at CERN North Area with a 
beam from the Super Proton Synchrotron (SPS) H4 beamline.
The ultrarelativistic electrons (positrons) of  $E_0 \simeq 100\, \mbox{GeV}$ can be exploited as the primary beam that is  scattering off nuclei of  an active thick target.  The typical scheme of the NA64 setup
can be found elsewhere in Ref.~\cite{NA64:2022rme}. 

The detector is equipped with:  (i) a low-material-budget tracker allowing the measurement of initial beam 
momentum with the precision of $1\%$; (ii) two magnetic spectrometers deflecting the primary beam line; (iii) 
synchrotron radiation detector (SRD) that is used for the  suppression of hadron beam contamination and the  effective tagging of the charged particles via their synchrotron radiation (SR); (iv)
the  active target represents electromagnetic calorimeter (ECAL) that is Shashlik-type 
modules consisting of alternating  plastic scintillator (Sc) and lead absorber (Pb). 

The fraction of the primary beam energy $E_{miss}=x E_0$
can be  carried away by  DM pair, that passes the NA64$e$ detector without energy deposition.
The remaining part of the beam energy fraction, $E_e^{rec} \simeq (1-x) E_0$, can be deposited in the 
electromagnetic  calorimeter (ECAL) of NA64$e$ by the recoil electrons (positrons).  
 For the NA64e experiment we use the following benchmark parameters 
 $(\rho~\simeq~11.34~~\mbox{g cm}^{-3}$, $A=207~~\mbox{g mole}^{-1}$, $Z=82$, 
 $X_0=2.56~~\mbox{cm})$, the effective interaction length of the electron is $L_T=X_0$ and the 
 missing energy fraction cut is $x_{min}=0.5$. The efficiencies $\eta_{\rm MED}^{\rm brem.}$ and 
 $\eta_{\rm MED}^{ann.}$ are taken to be at the level of $90\%$ for both electron and positron beam 
 modes~\cite{NA64:2022rme}.

 The 
 event selection rule for NA64e can be summarized as follows~\cite{NA64:2022yly}:
\begin{enumerate}
\item the beam track momentum  should be within $100\pm 3$ GeV; 
\item  the energy detected by the SRD should be consistent with the SR energy emitted by $e^\pm$'s in the magnets of the spectrometer;
\item  the shower shape  in the ECAL should be as expected from the signal-event  shower \cite{Gninenko:2016kpg} implying the cut for the recoil 
electron (positron)  $E^{rec}_e \lesssim 0.5 E_0 \simeq  50\,\mbox{GeV}$.
\end{enumerate}

  We note that about $\simeq 120$ days are needed to collect $\mbox{EOT} \simeq 5\times 10^{12}$ at 
 the $H4$ beam line for the projected statistics of the NA64$e$. In this work we also perform the 
 analysis of the sensitivity of NA64$e$ to probe DM for the 
 $\mbox{EOT} \simeq 3.22 \times 10^{11}$ (see e.~g.~Ref.~\cite{NA64:2022yly} for detail).

{ \it The light dark matter experiment (LDMX)} : is the projected electron fixed-target facility at Fermilab, that can be used for investigating the relic DM with the mass lying in the range between $1\, \mbox{MeV}$ and $1\, \mbox{GeV}$. The schematic layout of the LDMX experiment is given in Ref.~\cite{Berlin:2018bsc}. 
The projected LDMX facility would employ the aluminium target (Al) and the unique electron missing momentum 
technique~\cite{Mans:2017vej} that is complementary to the NA64$e$ facility. 
The  missing-momentum of the incoming beam can be measured by: (i) the beam tagger, (ii) target, (iii) silicon tracker system, (iv) electromagnetic and hadron calorimeter which are located downstream.
Thus, for the LDMX experiment we use the following benchmark parameters  $(\rho~=~2.7~~\mbox{g cm}^{-3}$, $A~=~27~~\mbox{g mole}^{-1}$, $Z~=~13$, $X_0~=~8.9~\mbox{cm})$ and $L_T ~\simeq~0.4~X_0~\simeq~3.56~~\mbox{cm}$. The typical efficiencies $\eta_{\rm MED}^{\rm brem.}$ and $\eta_{\rm MED}^{ann.}$ are estimated to be at the level of $\simeq 50 \%$ for both electron~\cite{Akesson:2022vza} and positron beam options.  

The energy of the primary beam is chosen to be $E_0\simeq16~~\mbox{GeV}$ and the projected moderate statistics 
corresponds to $\mbox{EOT}\simeq 10^{15}$ (it is planned however to collect $\mbox{EOT}\simeq 10^{16}$ by the final phase 
of experimental running after 2027, see e.~g.~Ref.~\cite{Akesson:2022vza} and references therein for detail). 
The rules for the missing momentum event selection in the LDMX are~\cite{Berlin:2018bsc}:
\begin{enumerate}
\item the beam track momentum  should be measured by the tagger at the level of $\simeq 16\,~\mbox{GeV}$; 
\item the silicon tracker  installed downstream the target should tag the large
transfer  momentum of the recoil particle $e^\pm$, that is associated with DM emission; 
\item the shower shape  in the ECAL should be as expected from the signal-event  shower \cite{Berlin:2018bsc} 
implying the cut for the recoil electron (positron)  $E^{rec}_e \lesssim 0.3  E_0 \simeq 4.8 \,\mbox{GeV}$.
\end{enumerate}
The explicit study on the 
background event estimate is provided in Ref.~\cite{LDMX:2019gvz} for the electron-bremsstrahlung beam mode only.  
We rely on that analysis and conservatively expect that for the positron beam mode of LDMX and the annihilation channel the background 
rejection would be the same.

\section{Typical thresholds
\label{SecThresholds}}

\begin{figure*}[!htb]
\centering
\includegraphics[width=0.49\textwidth]{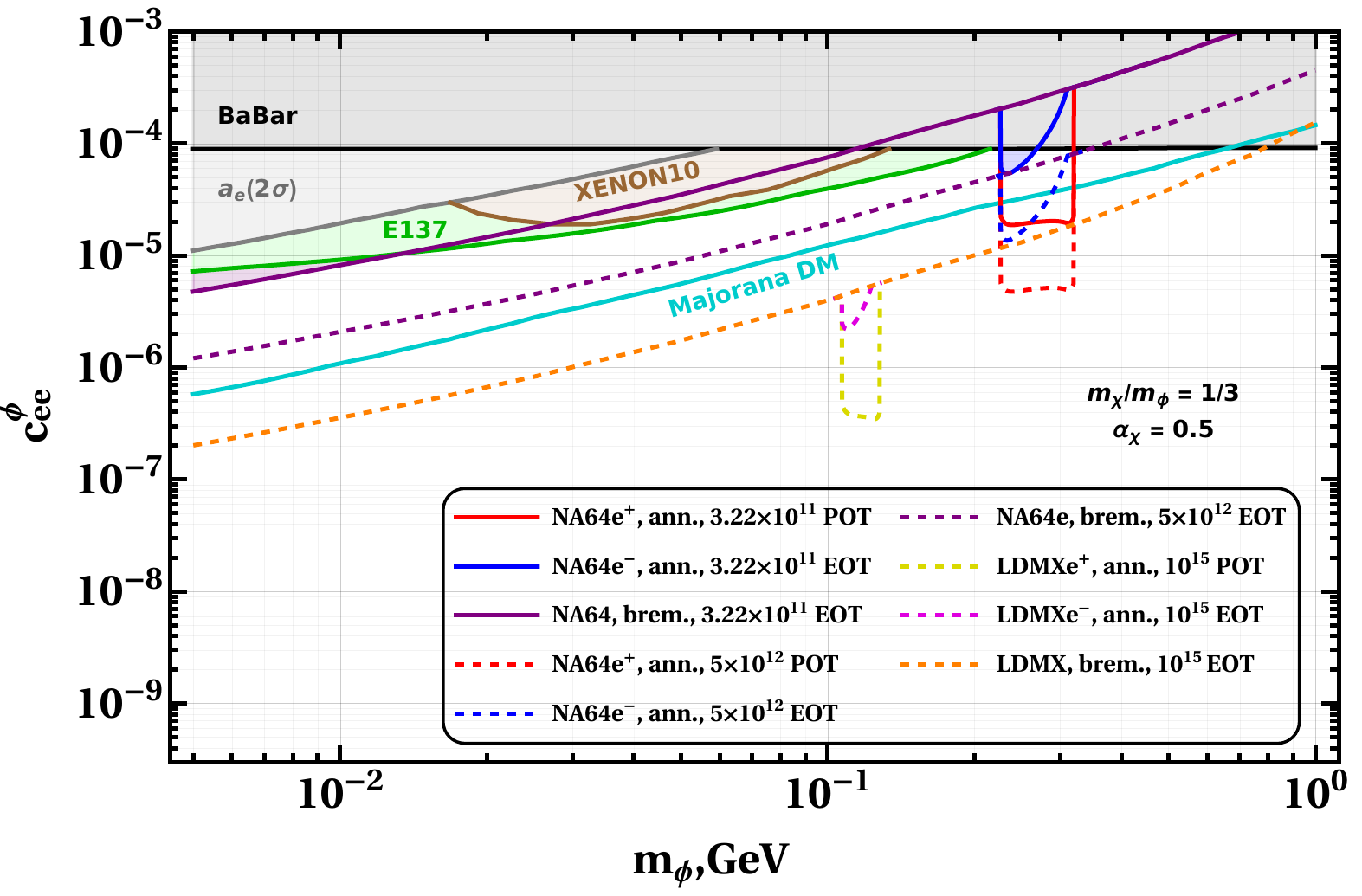}
\includegraphics[width=0.49\textwidth]{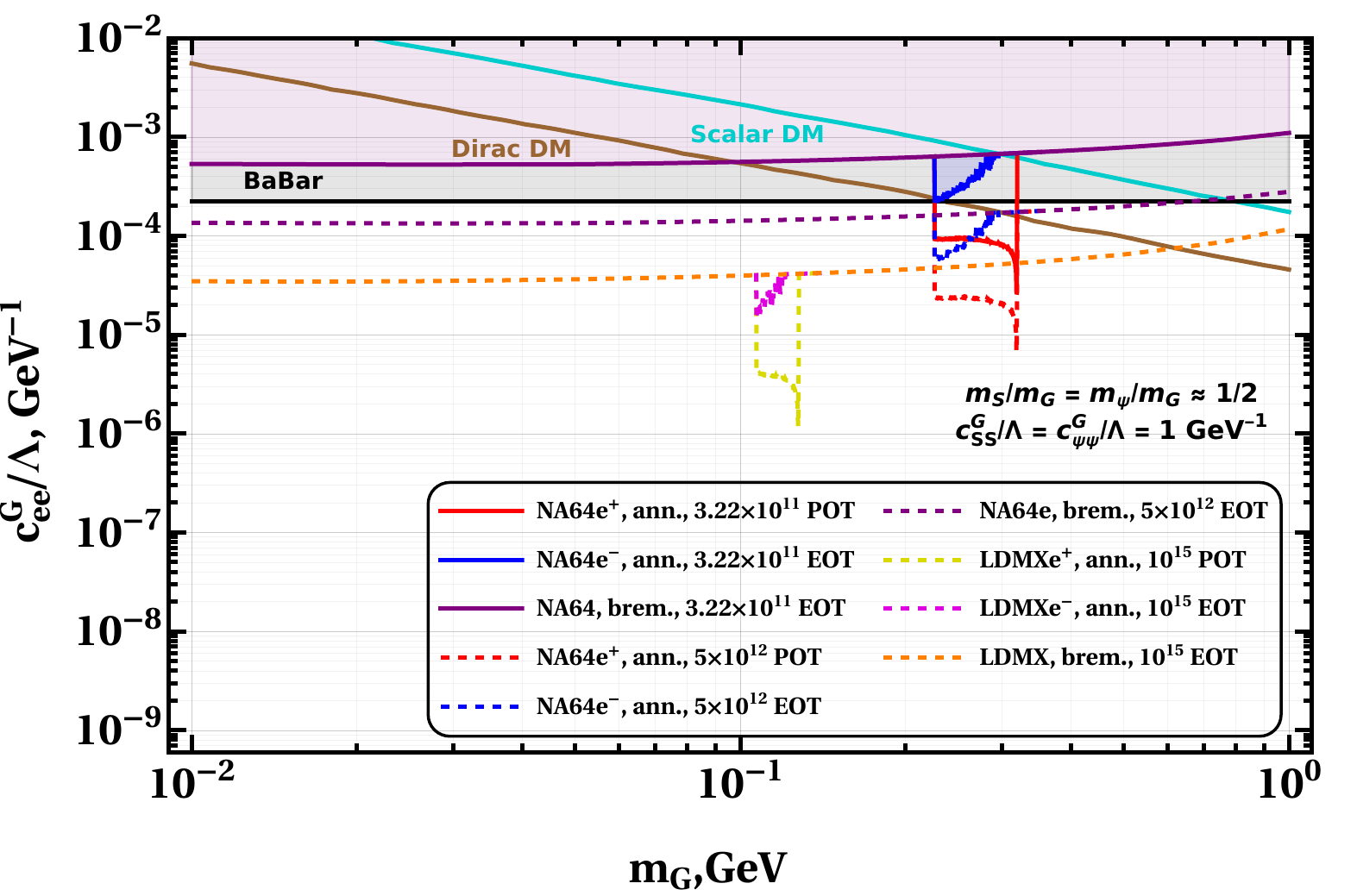}
\caption{Left panel: The experimental reach at $90\, \%$ C.L. for the NA64 and LDMX fixed target facilities
due to $\phi$-strahlung  and resonant positron annihilation $e^+e^- \to \phi$, followed by the invisible decay  
$\phi \to \chi \chi$. The solid purple line corresponds to the existing NA64 limits for $\mbox{EOT}\simeq 3.22\times 10^{11}$ 
from missing energy process $e N \to e N \phi$, the dashed purple line shows the  expected limit for projected 
statistics $\mbox{EOT}\simeq 5\times 10^{12}$.  the solid blue line is the existing limit of the NA64 
($\mbox{EOT}\simeq 3.22\times 10^{11}$) due to the positron  annihilation mode $e^+e^- \to \phi$ with primary $e^{-}$ 
beam,  the solid red line is the projected expected reach of NA64 for the positron annihilation channel 
with primary $e^+$ beam that corresponds to the $3.22\times 10^{11}$ positrons accumulated on target.
Dashed blue and red  lines correspond to the projected NA64 statistics $5\times10^{12}$ of the annihilation mode 
$e^+ e^- \to \phi$ for the $e^-$ and $e^+$ primary beam respectively. Dashed orange line is the projected LDMX
limits for $\mbox{EOT}\simeq  10^{15}$ from $\phi$-strahlung process $e N \to e N \phi$. 
Dashed pink and yellow  lines correspond to the projected LDMX statistics $10^{15}$ of the annihilation mode 
$e^+ e^- \to \phi$ for the $e^-$ and $e^+$ primary beam respectively. Grey shaded region corresponds to the exisiting 
BaBar~\cite{BaBar:2017tiz} monophoton limit $e^+e^- \to \gamma \phi$. Green shaded region is the current reach of the 
electron beam dump E137~\cite{Berlin:2018bsc,Bjorken:1988as,Batell:2014mga} experiment. 
Brown shaded region is the 
existing bound from  XENON10~\cite{XENON10:2011prx,Essig:2012yx,Essig:2017kqs} direct detection experiment. 
 Solid cyan line is thermal targets for Majorana dark matter that couples to an electrophilic spin-0 mediator $\phi$.
 The latter curve was adapted from top left panel in Fig.~9 of Ref.~\cite{Berlin:2018bsc}, it implies the 
 the benchmark DM  parameters $m_\chi/m_\phi=1/3$ and $\alpha_\chi \simeq 0.5$. All these curves imply that mediator 
 decays in the invisible mode, $\mbox{Br}(\phi\to \chi \bar{\chi})\simeq 1$. 
 Right panel: The same as in the left plot but 
 for the spin-2 mediator $G$. Shaded gray region corresponds to the current BaBar~\cite{Kang:2020-LGMDM} monophoton
 $e^+ e^- \to \gamma G$ reach. Solid brown and solid cyan lines corresponds to the thermal targets for Dirac DM and 
 scalar DM that couple to an electrophilic spin-2 mediator $G$. These curves were adapted from left panel in Fig.~8 of 
 Ref.~\cite{Kang:2020-LGMDM}, that implies the typical DM parameters 
 $c^{\rm G}_{SS}/\Lambda = c^{\rm G}_{\psi \psi}/\Lambda\simeq 1.0\, \mbox{GeV}^{-1}$ and 
 $m_S/m_G=m_\psi/m_G \simeq 0.498$. All these curves imply  $\mbox{Br}(G\to \psi \overline{\psi})\simeq 1$ 
 and $\mbox{Br}(G\to SS)\simeq 1$. It is worth noticing  again  that for these benchmarks, the DM bounds  are agnostic to the specific type  of DM, i.~e.~ the exclusion limits of scalar and Dirac DM 
 from NA64 (LDMX) are indistinguishable (see  text). 
}\label{fig:CoupConstr_hpi}
\end{figure*}

Let us consider now the kinematics of the above mentioned process in detail. 
As we discussed earlier, in case of fixed-target experiments, the process of the electron-positron annihilation 
arises form the interaction of secondary positrons from the EM shower  with
atomic electrons in the thick active target. The EM shower originates from the primary 
electron beam impinging on the fixed dump. We neglect the velocity of the  atomic electron, therefore
the typical momenta of the electrons and secondary positrons are chosen to be
 $p_{e^-} = (m_e, 0,0,0)$ and $p_{e^+} \simeq (E_{e^+}, 0, 0, E_{e^+})$, respectively.  
 Thus, the Mandelstam variable takes  the following form: 
\begin{equation}
    s  = (m_e + E_{e^+})^2 - |\vect{p}_{e^+}|^2 = m_e^2 + 2 m_e E_{e^+} \simeq 2 m_e E_{e^+}, 
\end{equation}
where we imply that positrons are ultrarelativistic and hence $|\vect{p}_{e^+}| \simeq E_{e^+}$.

In Fig.~\ref{fig:tot_cross_sec_fermion} for the illustrative purpose  we show  the resonant total cross sections (see 
e.~g.~Eqs.~(\ref{eq:tot_eeToGchichi}) and~(\ref{eq:tot_eeToSchichi}) for detail)
as a function of energy fraction between  secondary positrons and primary beam for the NA64e experiment. 

One can conclude from Fig.~\ref{fig:tot_cross_sec_fermion}, that the larger coupling 
of the mediator to DM, the large total cross section at the peak position in the case 
of invisible mode $\mbox{Br}(\mbox{MED}\to \mbox{DM}\; 
\mbox{DM})\simeq 1$.  However, the position of the  peak is related to the $m_{\rm MED}$ and   does not dependent on the 
typical  magnitude of  dark fine structure constant. The  contribution to the total cross section of
the process $e^+e^- \to \text{MED} \to \text{\rm DM}\;\text{\rm DM}$ is associated with  the off shell mediator
energy range $2 m_\text{\rm DM} \lesssim \sqrt{s} \lesssim m_{\text{MED}}$, and the typical 
resonant  masses  at  $m_{\rm MED}\simeq  \sqrt{2 m_e E_{e^+}}$.  As a result, one can estimate the mass range $m_{\rm MED}$
that would enhance the signal in the NA64 due to the MED emission. In particular, the  energy deposition cut for
NA64$e$ annihilation mode, $E_0/2 \lesssim E_{e^+} \lesssim E_0$, yields the following mass range 
$\sqrt{E_0 m_e}\lesssim m_{\rm MED} \lesssim \sqrt{2 E_0 m_e}$, that numerically corresponds to the typical signal mass bounds
$0.23 \, \mbox{GeV} \lesssim m_{\rm MED} \lesssim 0.32\, \mbox{GeV}$ for $E_0 \simeq 100\, \mbox{GeV}$. 

It is worth noticing that 
in the case of LDMX experiment, the shape of the annihilation cross section would be the same, however the typical 
resonant  mass range shifts to the 
$0.11 \,\text{GeV} \lesssim  m_{\rm MED} \lesssim  0.13 \, \text{GeV}$ which implies the missing energy 
cut  $0.7 E_0 \lesssim E_{e^+}\lesssim E_0$  for $E_0\simeq 16\, \mbox{GeV}$.
It turns out, that the resonant mass range of NA64e is $\simeq 4.48$ times
wider than the one for the LDMX facility.

One can estimate the typical  width of the cross section  peak
at the resonant point $E_{R} \simeq m_{\rm MED}^2/(2 m_e)$ in terms of positron energy $E_{e^+}$ for the NA64e facility.
The latter is crucial for the extraction of the signal yield in the narrow energy region. In 
particular, by solving the algebraic  equation for the Breit-Wigner term in the denominator 
($s-m_{\rm MED}^2 )^2~\simeq~m_{\rm MED}^2  \Gamma_{\rm MED}^2$ one finds 
$$
E_{\pm} \simeq E_R \pm \Delta E/2,
$$
where $\Delta E \simeq m_{\rm MED} \Gamma_{\rm MED}/m_e$.  
In the case of $0.23\, \mbox{GeV} \lesssim  m_{\rm MED}  \lesssim   0.32 \, \mbox{GeV}$, the range of energy width
corresponds to $ 0.037 \, \mbox{GeV} \lesssim \Delta E_{e^+} \lesssim  0.14 \, \mbox{GeV}$ for Dirac DM,  $\alpha_{\psi}\simeq 0.5 \, \mbox{GeV}^{-2}$
and $m_\psi/m_G \simeq 1/3$.

Note that for the scalar DM the  energy width  is fairly narrow 
and can be in the range $ 0.0026 \, \mbox{GeV} \lesssim \Delta E_{e^+} \lesssim  0.0098 \, \mbox{GeV}$ 
for $\alpha_S \simeq 0.5 \, \mbox{GeV}^{-2}$ and $m_S/m_G \simeq 1/3$. Finally, for the typical resonant
mass range of the LDMX facility $0.11\, \mbox{GeV} \lesssim  m_{\phi}  \lesssim   0.13 \, \mbox{GeV}$ the
positron energy resolution can be relatively large 
$ 1.22 \, \mbox{GeV} \lesssim \Delta E_{e^+} \lesssim  1.71 \, \mbox{GeV}$ for $\alpha_\chi\simeq 0.5$ 
and $m_\chi/m_\phi \simeq 1/3$.

 One can estimate the on shell mediator production cross section for the narrow-width regime
as long as  $\Gamma_{\rm MED} \to 0$ in the following form
\begin{align}
\label{eq:totDeltaG}
 &   \sigma_{e^-e^+ \to G \to \psi \overline{\psi}}
=
    \sigma_{e^-e^+ \to G \to S S}
=
    \delta(s - m_{G}^2)
    \widetilde{c}_G,
\\
&    \sigma_{e^-e^+ \to \phi \to \chi \chi}
=
    \delta(s - m_{G}^2)
    \widetilde{c}_\phi,
    \label{eq:totDeltaPhi}
\end{align}
where the dimensionless normalization prefactors read as follows:
\begin{equation}\label{eq:totWithoutDelta}
    \widetilde{c}_G = 5 \pi (c^G_{ee})^2 m_{G}^2 /(8 \Lambda^2), 
\qquad        \widetilde{c}_\phi = \pi (c^\phi_{ee})^2/2.
\end{equation}
In Eqs.~(\ref{eq:totDeltaG}) and~(\ref{eq:totDeltaPhi}) we use the representation of Dirac 
delta function:
\[
     \delta(x) = \lim_{\Gamma \to 0} \Gamma/[\pi (x^2 + \Gamma^2)].
\]
 It follows from Eqs.~(\ref{eq:totDeltaG}), (\ref{eq:totDeltaPhi}) and 
 (\ref{eq:totWithoutDelta}) that at the first order the resonant total cross section does not 
 depend on the coupling  between the DM and MED \cite{Andreev:2021fzd}. Moreover, for the the 
 spin-2 mediator the cross section is agnostic to the specific type of the outgoing DM 
 particles.

By taking into account that a dominant contribution to the resonant cross section is close to 
$s \simeq  m_{\rm MED}^2$ for the near-threshold approach $m_{\rm MED} \simeq  2 m_{\rm DM}$  $( \Gamma_{\rm MED} \to 0)$ we get the signal yield for the resonant process in the following form
\begin{multline}\label{eq:constrCoulConst}
    N^{\rm ann.}_{\rm MED} = \mbox{EOT} \times  \frac{N_A Z L_T \rho}{A} 
    \frac{\widetilde{c}_{\rm MED}  T_+(E_R)}{2 m_e}
\\ \! \times\!
   \theta \! \left(E_R - E^{cut}_{e^+} \right)
   \! \theta 
   \! \left(E^{max}_{e^+} - E_R \right),
\end{multline}
where $\widetilde{c}_{\rm MED}$ is  defined by Eq.~(\ref{eq:totWithoutDelta}), 
$E_R = m_{\rm MED}^2 / (2 m_e)$ is the typical energy of the resonance, $\theta(x)$ is the Heaviside step function.

\section{The experimental limits\label{SectionExpectedReach}}

In this section we study the experimental reach of fixed-target facilities NA64e and LDMX for both primary electron 
and positron  beams for spin-2 and spin-0 DM  
assuming their invisible decay mode.
For the background free case and the null results of the missing energy events associated with DM mediator production, we 
set the number of signal events  $N_{\rm sign.} \gtrsim  2.3$ and obtain the $90\% \mbox{~C.~L.}$ exclusion limit on the 
electron-specific mediator coupling constant. Here we suppose that the signal originates from the MED-strahlung and 
$e^+e^-$~annihilation mechanism,   
i.~e.~$N_{\rm sign. }\simeq N^{\rm brem. }_{\mbox{\scriptsize MED}}+N^{\rm ann. }_{\mbox{\scriptsize MED}}$.

In the left panel (right panel) of Fig.~\ref{fig:CoupConstr_hpi} we show the experimental reach of the NA64 and LDMX 
for the electron-specific scalar (tensor) mediator coupling  $c_{ee}^\phi$  ($c_{ee}^{\rm G}/\Lambda$). 
The expected limits associated with  $\phi$-strahlung ($G$-strahlung), $e N \to e N \phi (G)$, 
followed by invisible decay $\phi\to \chi \chi$ ($G\to \overline{\psi} \psi (SS)$)
are depicted by the dashed violet and dashed orange lines for NA64 and LDMX respectively. These 
expected reaches are derived for the projected statistics of NA64e at the level of $\mbox{EOT}\simeq 5\times 10^{12}$ and for the LDMX with $\mbox{EOT}\simeq 10^{15}$. 
The existing limit of NA64
on $c^{\phi}_{ee}$ is shown by the solid violet line, that implies
$\mbox{EOT}\simeq 3.22\times 10^{11}$. 

For the projected statistics associated with positron 
annihilation mode we consider  both positron  and electron  primary beam  options.
In Fig.~\ref{fig:CoupConstr_hpi} the dashed pink and yellow lines correspond to the LDMX expected reach associated with electron and 
position beam respectively. 

It follows from Fig.~\ref{fig:CoupConstr_hpi} that the positron beam gives more stringent constraint on the mediator coupling  constant due to the enhanced number of positron in the first generation of EM shower. In particular, for the LDMX mass range of interest $0.11 \,\text{GeV} \lesssim  m_{\rm MED} \lesssim  0.13 \, \text{GeV}$
the positron annihilation channel pushes down the exclusion limits by an order of magnitude.
Note that the LDMX can reach a fairly strong limits  on an electron-specific mediator due to the 
sufficiently large number of projected accumulated statistics  $\mbox{EOT}\simeq 10^{15}$. For instance, 
for the typical mass $ m_{\phi} =m_{G} \simeq 0.1\, \mbox{GeV}$ the LDMX is able to set the constraint 
at the level of $c^\phi_{ee}\simeq 5\times 10^{-7}$ and $c^{\rm G}_{ee}/\Lambda \simeq 10^{-6}\, \mbox{GeV}$ for  the spin-0 and spin-2 mediator  respectively.

It is worth noticing that the authors of Ref.~\cite{Berlin:2018bsc} have studied in detail
the thermal production mechanism of DM involving parity-even electron-specific spin-0 
mediator.   In  the left panel of Fig.~\ref{fig:CoupConstr_hpi} we show the  relic
abundance target lines for Majorana DM by the solid cyan line. That curve implies the benchmark DM 
parameters $m_\chi/m_\phi=1/3$ and $\alpha_\chi \simeq 0.5$.

Remarkably, for the projected statistics 
$\mbox{EOT}\simeq 5 \times 10^{12}$ the NA64 facility can probe the thermal Majorana DM parameter space   at 
the level of $c_{ee}^\phi \lesssim (3.0 - 4.0) \times 10^{-5}$ for the resonant channel $e^+e^-\to \chi \chi$  even  with a primary 
electron  beam mode (dashed blue line in left panel of Fig.~\ref{fig:CoupConstr_hpi}). The significant enhancement of the 
sensitivity $c_{ee}^\phi \lesssim 5.0 \times 10^{-6}$ can be achieved by exploiting the positron beam mode of NA64 for 
the typical masses in the range $0.23\, \mbox{GeV} \lesssim  m_{\rm MED}  \lesssim   0.32 \, \mbox{GeV}$  (dashed 
red line in left panel of Fig.~\ref{fig:CoupConstr_hpi}). 

It is important to note, that  the existing constraints of NA64 for $\mbox{EOT}\simeq 3.22 \times 10^{11}$ 
exceed the typical monophoton $e^+e^- \to \gamma \phi$ 
bound of the BaBar~\cite{BaBar:2017tiz} facility at the level of $c_{ee}^\phi \lesssim 6\times 10^{-5}$. 
Moreover, the current limits of NA64 associated with $\phi$-strahlung are complementary to the 
experimental reach of the E137~\cite{Berlin:2018bsc,Bjorken:1988as,Batell:2014mga} and 
XENON10~\cite{XENON10:2011prx,Essig:2012yx,Essig:2017kqs} direct detection experiment for 
$m_\phi \lesssim 10^{-2}\, \mbox{GeV}$  and $c_{ee}^\phi \lesssim 10^{-5}$.

In addition we note that  the thermal production mechanism of DM involving spin-2 electrophilic
mediator have been analyzed in detail in Ref.~\cite{Kang:2020-LGMDM}. To be more specific, in right panel of 
Fig.~\ref{fig:CoupConstr_hpi} we show the typical relic abundance target curves for the scalar (solid cyan line)
and Dirac DM (solid brown line) adapted from Ref.~\cite{Kang:2020-LGMDM}.  These lines imply the typical DM  
couplings $c^{\rm G}_{SS}/\Lambda = c^{\rm G}_{\psi \psi}/\Lambda\simeq 1.0\, \mbox{GeV}^{-1}$ 
and benchmark mass ratios that are close to the DM production threshold $m_S/m_G=m_\psi/m_G \simeq 0.498$.
This means that one can use the approximate Eq.~(\ref{eq:constrCoulConst}) in 
 the calculation of the yield of the spin-2 mediator, as long as $\Gamma_{G}\to 0$ for $m_{G}\simeq 2 m_{DM}$. 
 It is worth noticing  again  that for these benchmarks, the DM bounds of an electron fixed target 
 are agnostic to the specific type  of DM, i.~e.~ the exclusion limits of scalar and Dirac DM are 
 indistinguishable. 

The bound from invisible monophoton searches $e^+e^- \to \gamma G$  at BaBar applies to the  typical parameter space of  the spin-2 mediator with 
$m_G \lesssim 1\, \mbox{GeV}$ and $c_{ee}^{\rm G}/\Lambda \lesssim 2\times 10^{-4}\, \mbox{GeV}^{-1}$. This bounds rules out 
the existing $G$-strahlung constraints from NA64  for $\mbox{EOT}\simeq 3.22\times 10^{11}$. However, the resonant 
mechanism of DM production with NA64 (solid blue line in right panel of Fig.~\ref{fig:CoupConstr_hpi}) 
provides the bound on $c_{ee}^{\rm G}/\Lambda$ that barely touches the BaBar limit at 
$m_G\simeq 0.23\, \mbox{GeV}$. Remarkably, the latter constraint is associated also with Dirac DM relic curve. However, 
the scalar DM is almost ruled out by BaBar. We note that for the projected statistics $\mbox{EOT}\simeq 5\times 10^{12}$
of NA64 one can probe the Dirac DM at the level of $c_{ee}^{\rm G}/\Lambda\lesssim 2.0 \times 10^{-4}\, \mbox{GeV}^{-1}$
for $m_{G}\simeq 0.3 \, \mbox{GeV}$. Finally, we note that positron beam mode exploiting in the NA64  
(with $\simeq 5\times 10^{12}$ positrons accumulated on target)  can set a fairly strong bound at 
$c_{ee}^{\rm G}/\Lambda \lesssim 10^{-5}\, \mbox{GeV}^{-1}$ that is complementary to the  LDMX expected reach for 
$\mbox{EOT}\simeq 10^{15}$ at $m_G\simeq 3\times 10^{-1}\,~\mbox{GeV}$.  However  the latter facility can rule out the spin-2 parameter 
space $ c_{ee}^{\rm G}/\Lambda \lesssim 1 \times 10^{-6} \mbox{GeV}^{-1}$  in the resonant mode close to typical mass at
$m_G \simeq  1.5\times 10^{-1}\, \mbox{GeV}$.

\section{Conclusion
\label{SectionConclusion}}

In the present paper we  have studied in detail the resonant probing spin-0 and spin-2 
DM mediator with electron fixed target experiments NA64 and LDMX. In particular, we showed that $e^+$ resonant 
annihilation $e^+e^- \to \mbox{MED}\to \mbox{DM}+\mbox{DM}$ can be a viable mechanism of the 
electron-specific DM mediator production, such that this process is competitive with widely exploited 
MED-strahlung production, $eN\to e N \, \mbox{MED}$, followed by the invisible decay into pair of DM particles,
$\mbox{MED}\to \mbox{DM}+\mbox{DM}$. Moreover, we estimated the reach of NA64 and LDMX and showed that the 
exclusion limits are pushed down by factor of $\mathcal{O}(1)$ for the specific mass range of the mediators.

\begin{acknowledgments} 
We would like to thank A.~Celentano,  P.~Crivelli, S.~Demidov, R.~Dusaev,  S.~Gninenko, D.~Gorbunov,  
M.~Kirsanov, N.~Krasnikov, V.~Lyubovitskij, L.~Molina Bueno,  A.~Pukhov,  H.~Sieber, and 
A.~Zhevlakov    for very helpful discussions and  correspondences.
The work of D.~V.~K on description of the  dark matter missing energy signatures of NA64$e$ 
and  exclusion limits for spin-2 DM mediator is  supported
by the  Russian Science Foundation  RSF Grant No. 21-12-00379.

\end{acknowledgments}	

\appendix

\section{MAJORANA FERMION}\label{sc:MajorFerm}
In this section we calculate a width of decay and a resonant cross section for the spin-0 mediator with Majorana fermions in a final state. For calculations below, we use two-component spinors formalism from~\cite{Dreiner:2008tw}. 

The Lagrangian of interaction between spin-0 massive mediator $\phi$ and Majorana fermions $\chi$ is:
\begin{equation}
\mathcal{L}^{\phi}_{\rm Majorana \small } \supset - \frac{1}{2} c^\phi_{\chi\chi}  \phi \, \chi \, \chi + \frac{1}{2} m_{\chi} \chi \, \chi + \mbox{H.~c.}
\end{equation}
The matrix element for a decay of spin-0 mediator $\phi$ into pair of Majorana fermions $\phi(p, s)~\to~\chi(p_1, s_1) \chi(p_2, s_2)$ takes  the following form~\cite{Dreiner:2008tw}:
\begin{equation}
i M = - i c^\phi_{\chi\chi} (y_1  y_2 + {x^\dagger_1} {x^\dagger_2} ),
\end{equation}
where  the two-component spinors are
\[
    {x_i}^{\alpha} = x^{\alpha}(\vect{p}_i,s_i), 
\quad
    {y_i}^{\alpha} = y^{\alpha}(\vect{p}_i,s_i).
\]
The summation over spinor indices  is defined as  
$
    {y_1}^{\alpha} {\delta_{\alpha}}^{\beta} {y_2}_{\beta} = y_1 y_2.
$
Moreover we imply the notation $x^{\alpha}y_{\alpha}$ and $x_{\dot{\alpha}}y^{\dot{\alpha}}$, 
where the undotted and dotted indices  correspond  to the left-handed and right-handed fermions, 
respectively, they are related via the Hermitian conjugation. 

Also, for a raising and a lowering  indices  we use the antisymmetric tensor $2 \times 2$ that has the following form: 
\[
    \epsilon^{12}~=~-~\epsilon_{12}~=~1, 
\qquad
    \epsilon_{\alpha \beta} \epsilon^{\gamma \delta} 
=
    \delta^{\delta}_{\alpha} \delta^{\gamma}_{\beta}
-
    \delta^{\gamma}_{\alpha} \delta^{\delta}_{\beta}.
\]
The (anti)~commutation relations read \cite{Schwartz:2014sze}:
\[
    x y = x^{\alpha}y_{\alpha} = - x_{\alpha}y^{\alpha} = y^{\alpha} x_{\alpha} = y x, 
\quad
    x y = - y x.
\]
By taking into account the spin sum \cite{Dreiner:2008tw}, we get expressions in the
following form:
\begin{multline*}
    \sum\limits_{spins}
    ( y_1 y_2 ) ( y^\dagger_2 y^\dagger_1 )
=
    \sum\limits_{spins}
    {\delta_{\alpha}}^{\beta} 
    {\delta_{\dot{\gamma}}}^{\dot{\lambda}}
    {y_1}^{\alpha} 
    {y_2}_{\beta}
    {y_2^\dagger}_{\dot{\lambda}}
    {y_1^\dagger}^{\dot{\gamma}} 
= \\ =
    ( p_2, \sigma_{\beta \dot{\lambda}} )
    ( p_1, \overline{\sigma}^{ \dot{\lambda} \beta } )
=
     {p_2}^\mu {p_1}^\nu
     \text{Tr} (\sigma_\mu \overline{\sigma}_\nu)
=
     2 ( p_1 , p_2 ),
\end{multline*}
\begin{multline*}
    \sum\limits_{spins}
    x_1^\dagger x_2^\dagger
    y_2^\dagger y_1^\dagger
=
    \sum\limits_{spins}
    {\delta_{\dot{\alpha}}}^{\dot{\beta}}
    {\delta_{\dot{\gamma}}}^{\dot{\lambda}}
    {x_1^\dagger}_{\dot{\beta}}
    {x_2^\dagger}^{\dot{\alpha}}
    {y_2^\dagger}_{\dot{\lambda}}
    {y_1^{\dot{\gamma}\dagger}}
= \\ =
    - m_\chi^2
    {\delta_{\dot{\alpha}}}^{\dot{\beta}}
    {\delta_{\dot{\gamma}}}^{\dot{\lambda}}
    {\delta_{\dot{\lambda}}}^{\dot{\alpha}}
    {\delta_{\dot{\beta}}}^{\dot{\gamma}}
=
    - m^2
    {\delta_{\dot{\beta}}}^{\dot{\beta}}  
=
    -2 m_\chi^2,
\end{multline*}
\begin{equation*}
    \sum\limits_{spins}
    y_1 y_2 
    x_2 x_1 
=
    -2 m_\chi^2,
\quad
    \sum\limits_{spins}
    x_1^\dagger x_2^\dagger 
    x_2 x_1 
=
    2 ( p_1 , p_2 ),
\end{equation*}
where  $\sigma^{\mu} = (\mathbb{I}_{2\times2},\vect{\sigma})$, $\overline{\sigma}^{\mu} = (\mathbb{I}_{2\times2}, - \vect{\sigma})$ with $\vect{\sigma}$ being a matrices of Pauli.
Finally, the   matrix element squared by assuming the summation over
the final states is:
\begin{equation}
    \sum\limits_{spins}
    | M |^2 = 
    4 (c^\phi_{\chi\chi})^2
    \left[
        (p_1, p_2) - m_\chi^2
    \right].
\end{equation}
The   decay width can be written as follows
\begin{equation}
    \Gamma_{\phi\to \chi \chi} =
 \frac{1}{2}  \cdot    \frac{1}{16 \pi m_\phi^3} \lambda^{1/2}(m_\phi^2, m_{\chi}^2, m_\chi^2)  \sum\limits_{spins}
    | M |^2,
    \label{GammaTotMajAppendix}
\end{equation}
where $\lambda(x,y,y) = x^2(1-4y/x)$ is the triangle function, the extra-factor of $1/2$ in Eq.~(\ref{GammaTotMajAppendix}) 
is due to the identical Majorana particles in the final state~\cite{Dreiner:2008tw}. As a result one has 
\begin{equation}
  \Gamma_{\phi\to \chi \chi} = \frac{1}{4} \alpha_\chi m_\phi (1-4m_\chi^2/m_\phi^2)^{3/2}.
 \end{equation}
The cross section reads 
\begin{multline}\label{eq:tot_eeToSchichiMajorana}
    \sigma_{e^-e^+ \to \phi \to \chi \chi} \!\!
= \!  \frac{1}{2} \cdot 
    \frac{4 \pi \alpha_\chi (c^{\phi}_{ee})^2}{16 \pi }
    \frac{s 
        \left( 1 - \frac{4 m_\chi^2}{s}\right)^{3/2}
        }{(s - m_{\phi}^2)^2 + m_{\phi}^2 \Gamma_{\phi}^2}.
\end{multline}

\newpage
\bibliography{bibl}

\end{document}